%Paper: hep-th/9307185
%From: eguchi%tkyvax.hepnet@Lbl.Gov
%Date: Thu, 29 Jul 93 23:32:38 PDT

\input phyzzx

\date{July, 1993}
\rightline{University of Tokyo preprint UT 650}
\titlepage
\vskip 3cm
\title{c=1 Liouville Theory Perturbed by the Black-Hole Mass
Operator}
\author
{Tohru Eguchi}
\address
{Department of Physics, Faculty of Science}
\address
{University of Tokyo}
\address
{Tokyo, Japan 113}

\abstract{
We discuss the properties of the Liouville theory coupled to the
$c=1$ matter when perturbed
by an operator, the screening operator of the $SL(2;R)$ current
algebra,
which is supposed to generate the mass of
the two-dimensional black hole.
Mimicking the standard
KPZ scaling theory of the Liouville system perturbed by the
cosmological constant operator,
we develop a scaling theory of
correlation functions as functions of the mass of the black hole.
Contrary to the case of KPZ, the present theory does not have the
$c=1$ barrior and seems somewhat insensitive to the details of the
matter content of the theory; the string succeptibility equals
$1$ independent of the matter central charge.
It turns out that our scaling exponents agree with those of
the deformed matrix model recently proposed by Jevicki and Yoneya.}
\endpage
\overfullrule=0pt

%%%%%%%%%%%%%%%%%%%%% References
\def\cmp#1{{\it Comm. Math. Phys.} {\bf #1}}
\def\pl#1{{\it Phys. Lett.} {\bf B#1}}

\def\np#1{{\it Nucl. Phys.} {\bf B#1}}

\def\mpl#1{{\it Mod. Phys. Lett.} {\bf A#1}}
\def\prd#1{{\it Phys. Rev.} {\bf D#1}}
%%%%%%%%%%%%%%%%%%%%% References
\REF\BK{M. Bershadsky and D. Kutasov, \pl{266} (1991) 345.}
\REF\WITA{E. Witten, \prd{44} (1991) 314.}
\REF\EKY{T. Eguchi, H. Kanno and S.K. Yang, \pl{298} (1993) 73.}
\REF\WITB{E. Witten, \np{373} (1992) 187.}
\REF\TATA{G. Mandal, A. Sengupta and S. Wadia, \mpl{6} (1991) 1685;
I. Bars and D. Nemeschansky, \np{348} (1991) 89;
S. Elizur, A. Forge and E. Rabinovici, \np{359} (1991) 581.}
\REF\KPZ{V. Knizhnik, A. Polyakov and A. Zamolodchikov, \mpl{3}
(1988) 819.}
\REF\DDK{F. David, \mpl{3} (1988) 1651; J. Distler and H. Kawai,
\np{321} (1989) 509.}
\REF\LZ{B. Lian and G. Zuckerman, \pl{266} (1991) 21.}
\REF\BMP{P. Bouwknegt, J. McCarthy and K. Pilch, \cmp{145}
(1992) 541.}
\REF\KP{I. Klebanov and A. Polyakov, \mpl{6} (1991) 3237.}
\REF\K{I. Kostov, \pl{215} (1988) 499.}
\REF\GKN{D. Gross, I. Klebanov and M. Newman, \np{350} (1991) 621.}
\REF\KITA{Y. Kitazawa, \pl{265} (1991) 262; VI. S. Dotsenko,
\mpl{6} (1991) 3601; N. Sakai and Y. Tanii, "Correlation Functions
of c=1 Matter coupled to Two-Dimensional Gravity" ,
Tokyo Institute of Technology preprint TIT/HEP-168, March 1991.}
\REF\DVV{R. Dijkgraaf, H. Verlinde and E. Verlinde, \np{371} (1992)
269.}
\REF\CM{S. Chaudhuri and D. Minic, "On the Black Hole Background of
Two-Dimensional String Theory", Univ. of Texas preprint,
UTTG-31-92, December 1992; S.K. Rama, "Tachyon Back Reaction on d=2
Black Hole", Trinity College Dublin preprint, TCD-3-93,
March 1993.}
\REF\MO{N. Marcus and Y. Oz, "The Spectrum of the 2D Black Hole
or Does the 2D Black Hole has Tachyonic or W Hair?",
Tel-Aviv University preprint, TAUP-2046-93, May 1993.}
\REF\CH{S. Chandrasekhar, {\it The Mathematical Theory of Black Holes},
Clarendon Press, Oxford, England 1983.}
\REF\JY{A. Jevicki and T. Yoneya, "A Deformed Matrix Model and the
Black
Hole Background in Two-Dimensional String Theory", Brown preprint
BROWN-HEP-904, May 1993.}
\REF\MS{E. Martinec and S. Shatashvili, \np{368} (1992) 338.}
\REF\DAS{S. Das, \mpl{8} (1993) 69; A. Dhar, G. Mandal and S. Wadia,
\mpl{7} (1992) 3703.}
\REF\D{Ulf Danielsson, "A Matrix-Model Black Hole", CERN preprint,
CERN-TH 6916/93, June 1993; K. Demeterfi and J. Rodrigues,
"States and Quantum Effects
in the Collective Field Theroy of a Deformed Matrix Model",
Princeton preprint PUPT-1407, June 1993.}
\REF\KU{S. Kachru, \mpl{7} (1992) 1419; J. Barbon, "Perturbing the
Ground
Ring of 2-D String Theory", CERN preprint,
CERN-TH 6379/92, January 1992.}

%%%%%%%%%%%%%%%%%%%%%%%%%%%%%%%%%%%%%%%%%%%%%%%%%%%%%%%%%%%%%%%%%%

In this article we would like to discuss the behavior of the Liouville
theory coupled to the $c=1$ matter when it is perturbed by an operator
$V$ which generates a mass for the two-dimensional black
hole. The black-hole mass operator is essentially the screening
operator of the
$SL(2;R)$ current algebra [\BK] and has appeared in the
study of the two-dimensional black hole
based on the $SL(2;R)/U(1)$ gauged WZW model [\WITA].
If one uses the operator correspodence between the $c=1$
Liouville and $SL(2;R)/U(1)$ coset theories [\EKY], the operator
$V$ is expressed as
$$
V=(\partial X-i\sqrt{k' \over k}\partial\phi)\exp(-\sqrt{2 \over k'}
\phi)
\eqno\eq
$$
where $\phi$ and $X$ are the Liouville and matter field, respectively
and the $k=9/4$ is the level of $SL(2;R)$ current algebra
($k'=k-2=1/4$).
Up to a total derivative
the operator $V$ coincides with the $j=1,m=0$ member of the
negative (anti-Seiberg) discrete state of the $c=1$ Liouville theory
, $W^{-}_{1,0}=\partial X \exp(-2\sqrt{2}\phi)$ (we use the same
notations as in ref.[\WITB]).
It is easy to see that the action of the Liouville theory
perturbed by the black-hole mass term
$$
\eqalign{
&S_{M}={1 \over 8\pi}\int d^2x \sqrt{\hat{g}}\{
\hat{g}^{ab}(\partial_a\phi
\partial_b\phi+\partial_a X\partial_bX)-Q\hat{R}^{(2)}\phi \cr
&-\sqrt{{k' \over2}}M\hat{g}^{ab}(\partial_a X-i\sqrt{k' \over k}
\partial_a\phi)
(\partial_b X-i\sqrt{k'\over k}\partial_b\phi)
\exp(-\sqrt{2 \over k'}\phi) \},~~Q=\sqrt{2 \over k'} \cr}
\eqno\eq
$$
has the familiar form of the two-dimensional black hole [\TATA,\WITA]
$$
S={k \over 4\pi}\int d^2x \sqrt{\hat{g}}\hat{g}^{ab}
(\partial_a r\partial_b r+tanh^2 r
\partial_a\theta\partial_b\theta)
-{1 \over 4\pi} \int d^2x \sqrt{\hat{g}} \hat{R}^{(2)}\log cosh r
\eqno\eq
$$
after a change of variables
$$
\phi=\sqrt{2k'}\log cosh r + \phi^*~,~~X=\sqrt{2k}(\theta-
i\log tanh r)
\eqno\eq
$$
in the leading order of $1/k$. In eq.(2) $M$ denotes the mass of
the black hole and
is related to the value of the dilaton field $\Phi=\sqrt{{2 \over k'}}
\phi$ at the horizon $r=0$ [\WITA],
$$
M=\sqrt{{2 \over k'}}\exp(\Phi(r=0))=\sqrt{{2 \over k'}}
\exp(\sqrt{{2 \over k'}}\phi^*)~ .
\eqno\eq
$$

It turns out that the mass-perturbed Liouville theory eq.(2)
has properties which are very different
from those of the Liouville
system perturbed by the cosmological constant operator.
The action of the latter theory is given by
$$
\eqalign{
&S_{\mu}={1 \over 8\pi}\int d^2x \sqrt{\hat{g}}\{
\hat{g}^{ab}(\partial_a\phi
\partial_b\phi+\partial_a X\partial_b X)
-2i\alpha_{0}\hat{R}^{(2)}X
-Q\hat{R}^{(2)}\phi \cr
&+\mu W^{-}_{0,0}\bar{W}^{-}_{0,0} \}~,~~~~~~~Q=\sqrt{{25-c \over 3}}~.
\cr}
\eqno\eq
$$
In (6) the Liouville field is coupled to the matter with a central
charge
$c=1-12\alpha_{0}^2$ and the cosmological constant operator is given by
$$
W^{-}_{0,0}=W^{+}_{0,0}=\exp\alpha\phi,
{}~~~\alpha={-\sqrt{25-c}+\sqrt{1-c} \over \sqrt{12}}~.
\eqno\eq
$$
As is well-known, the two-dimensional gravity coupled to matter fields
(6) exists only
for $c\leq1$: the exponent
$\alpha$ becomes complex for $c>1$ and the theory does not make sense
above the $c=1$ barrior.

On the other hand, our theory (2) does not seem to have the barrior
at $c=1$ and exists for all values of $c<25$. In the case of
general values of the central charge eq.(2) is replaced by
$$
\eqalign{
&S_{M}={1 \over 8\pi}\int d^2x \sqrt{\hat{g}}\{
\hat{g}^{ab}(\partial_a\phi
\partial_b\phi+\partial_a X\partial_bX)
-2i\alpha_0\hat{R}^{(2)}X-Q\hat{R}^{(2)}\phi \cr
&-\sqrt{{k' \over2}}M\hat{g}^{ab}(\partial_a X-i\sqrt{k'\over k}
\partial_a\phi)
(\partial_b X-i\sqrt{k' \over k}\partial_b\phi)
\exp(-\sqrt{2 \over k'}\phi) \},~~Q=\sqrt{2 \over k'} \cr}
\eqno\eq
$$
where $k=2(28-c)/(25-c), k'=k-2$ and $c=1-12\alpha_0^2$.
As we see in eq.(8) the
exponent of our perturbation
operator is equal to (minus) the background charge $Q$
for all values of $c$ and is well-defined up to
$c<25$.

We can in fact mimick the KPZ scaling theory [\KPZ] of the
two-dimensional
gravity (6)
and develop a similar
scaling analysis of our system (8). As one can check easily,
critical
exponents of the theory eq.(8) are all well-defined for $c<25$
and do not show
pathologies at $c>1$.

Let us first consider the partition funciton of (8)
$$
Z(M)=\int{\cal D}\phi {\cal D}X \exp(-S_{M}(X,\phi))
\eqno\eq
$$
(we suppress contributions from the ghost fields)
and assume its power-behavior in the parameter $M$
$$
Z(M)\cong M^{2-\gamma_{M}}~.
\eqno\eq
$$
Eq.(10) defines the string succeptibility $\gamma_{M}$
of our system.
In order to compute $\gamma_{M}$
we use the heuristic method of derivation of
the KPZ scaling [\DDK]: one shifts the Liouville field by a constant
$\phi \rightarrow \phi+\rho$ in eq.(9) and uses the Gauss-Bonnet
formula
$1/4\pi\int d^2 x \sqrt{\hat{g}}\hat{R}^{(2)}=2-2g$
($g$ is the genus of the
world-sheet) to obtain,
$$
Z(M)=e^{Q\rho(2-2g)/2} Z(Me^{-Q\rho})~.
\eqno\eq
$$
We put $g=0$ hereafter.
Then the string succeptibility of our model equals
$$
\gamma_{M}=1
\eqno\eq
$$
and is independent of the matter central charge.

Let us recall that in case of the two-dimensional gravity
$$
\eqalign{
&Z(\mu)=\int {\cal D}\phi{\cal D}X \exp(-S_{\mu}(X,\phi)) \cr
&\cong \mu^{2-\gamma_{\mu}} \cr}
\eqno\eq
$$
the string succeptibility is given by
$$
\gamma_{\mu}=-{1 \over p}
\eqno\eq
$$
for the Liouville theory coupled to the $(p,p+1)$
unitary minimal matter with
$c=1-6/p(p+1)$.
$\gamma_{\mu}$ of (14) is negative for $c<1$ and vanishes at $c=1$.
By comparing (12) and (14) we see that
our system and
two-dimensional gravity have very different characteristics.
(At $c=1$ there exists a logarithmic scaling violation and $Z(\mu)$
behaves as $Z(\mu)\cong \mu^2\log\mu$.
We also expect a logarithmic scaling
violation at $c=1$ in the mass-perturbed theory and
$Z(M)\cong M\log M$.
Logarithmic terms are not detected by the simple method of
shifting the Liouville field).

Let us next turn to the discussion on correlation functions and their
scaling relations.
We first consider the case $c<1$ and recall the KPZ scaling
exponents for the gravitationally-dressed primary fields
of the $(p,p+1)$ minimal unitary series [\KPZ],
$$
\eqalign{
&\langle\Phi_{r,s}\rangle_{\mu}\equiv Z(\mu)^{-1}\int{\cal D}
\phi{\cal D}X\int d^2 x \sqrt{\hat{g}}\Phi_{r,s}(x)
\exp(\beta_{r,s}\phi(x))\exp(-S_{\mu}(X,\phi)) \cr
&\cong \mu^{-1+\Delta_{r,s}}~, \cr
&\Delta_{r,s}=1-{\beta_{r,s} \over \alpha}={r(p+1)-sp-1 \over 2p}~,
{}~~~~1\leq s\leq r\leq p-1~.
\cr}
\eqno\eq
$$

On the other hand, our theory predicts
$$
\eqalign{
&\langle\Phi_{r,s}\rangle_{M}\equiv Z(M)^{-1}\int{\cal D}
\phi{\cal D}X \int d^2 x \sqrt{\hat{g}}\Phi_{r,s}(x)
\exp(\beta_{r,s}\phi(x))\exp(-S_{M}(X,\phi)) \cr
&\cong M^{-1+\Delta_{r,s}}~, \cr
&\Delta_{r,s}=1+{\beta_{r,s} \over Q}
={1 \over 2}+{r(p+1)-sp \over 2(2p+1)}~,~~~~1\leq s\leq r\leq p-1~.
\cr}
\eqno\eq
$$
Thus the cosmological costant operator $(r=s=1)$ now has a scaling
exponent $\Delta_{1,1}=(p+1)/(2p+1)$.

Let us next turn to the case of $c=1$.
BRST invariant observables of the $c=1$ Liouville theory are
well-known [\LZ,\BMP,\WITB,\KP].
There exist tachyon states given by
$$
\eqalign{
&T^+_p(z)=\exp(i\sqrt{2}pX(z))\exp(\sqrt{2}(-1+\vert p \vert)\phi(z))~, \cr
&T^-_p(z)=\exp(i\sqrt{2}pX(z))\exp(\sqrt{2}(-1-\vert p \vert)\phi(z))~
\cr}
\eqno\eq
$$
where the Euclidean momentum $p$ takes arbitrary real values.
There also exists discrete states which
occur at special values of the momonta. $W_{\infty}$ currents,
for instance, are given by
$$
\eqalign{
&W^{\pm}_{j,j}(z)=T^{\pm}_j(z)~,
{}~~~~~~~~~~~~~~~~~~~~~~~~~~~~~~~~~~~~~j=0,1/2,1,\cdots \cr
&W^{\pm}_{j,m}(z)=(\oint\exp(-i\sqrt{2}X(w))dw)^{j-m}W^{\pm}_{j,j}(z)~,~~~
-j\leq m\leq j~~.
\cr}
\eqno\eq
$$
Let us first consider the tachyon 2-point function,
$$
G(M)=\langle T^+_{-p}T^+_p\rangle_{M}
\equiv\int{\cal D}\phi{\cal D}X T^+_{-p}T^+_p\exp(-S_M(X,\phi))
\eqno\eq
$$
where
$$
T^+_{p}=\int d^2 x\sqrt{\hat{g}(x)}\,T^+_{p}(x)~.
\eqno\eq
$$
Again by shifting the Liouville field $\phi \rightarrow \phi+\rho$
we obtain
$$
G(M)=e^{2\sqrt{2}(-1+\mid p \mid)\rho}e^{2\sqrt{2}\rho}
G(Me^{-2\sqrt{2}\rho})
\eqno\eq
$$
Hence
$$
G(M)=c_M(p)M^{\mid p \mid}~.
\eqno\eq
$$
$c_M(p)$ is some function depending only on the momentum $p$. (22) is
to be contrasted
with the case of the $c=1$ two-dimensional gravity [\K,\GKN,\KITA]
$$
\eqalign{
&G(\mu)=\langle T^+_{-p}
T^+_p\rangle_{\mu}\equiv\int{\cal D}\phi{\cal D}X T^+_{-p}
T^+_p\exp(-S_{\mu}(X,\phi)) \cr
&=c_{\mu}(p)\mu^{2\mid p \mid}~.
\cr}
\eqno\eq
$$
The normalization factor $c_{\mu}(p)$ in the right-hand-side of (23)
is known to have (double) poles at $\vert p \vert=$half-integers
[\GKN]
$$c_{\mu}(p)\sim (\Gamma(1-2\vert p \vert))^2
\eqno\eq
$$
In the present case it is somewhat difficult to compute $c_M(p)$
in a closed form due to the presence of the derivative terms $\partial
\phi,\partial X$ in the perturbing operator. It is, however, possible
to check that $c_M(p)$ has a double pole at $\vert p \vert=1$ but it is
regular at $\vert p \vert=1/2$. It appears quite likely
that $c_M(p)$ has the
following pole structure
$$
c_M(p)\sim(\Gamma(1-\vert p \vert))^2~.
\eqno\eq
$$

Appearance of poles in the 2-point function at special values of the
momenta indicates some resonance phenomenon.
We would like to point out that the behavior (25) suggests the
existence of the ringing or quasi-normal modes in two-dimensional
black hole which are quite analoguous to those known in the
four-dimensional black holes. In order to make further
analysis let us now switch to the target-space description of the
tachyon field and consider its propagation equation in the background
of the two-dimensional black hole
$$
\nabla^2 T -2\nabla\Phi\nabla T+2T=0~.
\eqno\eq
$$
Under a suitable choice of coordinates eq.(26) is reduced to a
hypergeometric equation and one can easily construct its solutions
[\WITA,\DVV,\CM,\MO]. One considers a solution of (26)
which is regular at the horizon and propagates towards the
center of the black hole
and decomposes it into a sum of ougoing and incoming waves at
asymptotic null-infinities. After simple calculations one finds
$$
T_p={1 \over \pi}\Gamma(1+\vert p \vert)\Gamma(-\vert p \vert)
e^{i\sqrt{2}pt}e^{-\sqrt{2}(1+\vert p \vert)r}+
{\Gamma(1+\vert p \vert)\Gamma(\vert p \vert)
\over \Gamma(1/2+\vert p \vert)^2}
e^{i\sqrt{2}pt}e^{-\sqrt{2}(1-\vert p \vert)r}.
\eqno\eq
$$
The first (second) term in the right-hand-side of eq.(27) represents an
outgoing (incoming) wave at infinity. If we renormalize the tachyon
wave-function by the factor $\Gamma(1-\vert p \vert)$ as suggested by
eq.(25), the coefficient of the
incoming wave of (27) vanishes at (Euclidean momoenta)
$\vert p \vert$=positive integers. Eq.(27) is then intepreted as
describing the decay of a
resonance into waves which either fall into the black hole or
escape to asymptotic infinity.
This phenomenon is quite analogous to the quasi-normal modes or
ringing modes known in the four-dimensional black hole theory (see for
instance, [\CH]).
Quasi-normal modes are defined by the expansion coefficients of a
solution of the wave equation propagating towards the interior of the
black hole at the horizon into a sum of
waves at asymptotic infinity
as in eq.(27),
$$
T(\omega)=a(\omega)e^{i\omega t-i\omega r}+b(\omega)
e^{i\omega t+i\omega r} .
\eqno\eq
$$
The zero of the coefficient function $b(\omega)$ in the upper-half
complex $\omega$ plane gives the characteristic frequency of a decaying
resonant state. It is known that in a large class of perturbations and
black hole geometries there exist an infinite number of quasi-normal
modes [\CH].
The peculiarity of our two-dimensional case is that the quasi-normal
frequencies occur all at pure imaginary values. It will be very
interesting
to check if other discrete states $\{W^{\pm}_{j,m}\}$ besides the
discrete tachyons $\{W^+_{j,j},j=1,2,\cdots\}$ are interpreted as
ringing modes of the two-dimensional black hole.

It is well-known that the two-dimensional gravity coupled to matter
fields (6) has an
alternative, discrete formulation by means of the large-$N$ matrix
models. In particular the $c=1$ case may be described by the matrix
quantum mechanics
$$
H={1 \over 2}Tr\big({d\Phi(t) \over dt} \big)^2-{1 \over 2}Tr\Phi(t)^2
\eqno\eq
$$
where $\Phi(t)$ is an $N\times N$ hermitian matrix
and the theory (29) is solved by converting it into that of
free fermions.
We wonder if there exists a similar discrete formulation for our
mass-perturbed $c=1$ Liouville system.

Recently a new type of matrix model, the deformed matrix model,
has been introduced by Jevicki and Yoneya
[\JY] in order to discuss
black holes within the framework of the matrix model.
The deformed model is defined by the Hamiltonian with an
inverse-squared potential
$$
H={1 \over 2}Tr\big({d\Phi(t) \over dt} \big)^2-{1 \over 2}Tr\Phi(t)^2
+{M \over N^2}Tr\Phi(t)^{-2}~,
\eqno\eq
$$
where the parameter $M$ is to identified as the mass of the black hole.
Contrary to the treatment of the model (29) where the system approaches
criticality as the fermi energy reaches its critical value,
$\mu_{cr}-\mu_{F}=\mu\rightarrow 0$,
one puts $\mu=0$ in the deformed model (30) and drives the system to a
crtitical point
by taking the strength of the repulsive potential go to zero,
$M\rightarrow 0$.
Unfortunately, it is not straightforward to interprete the model (30)
as describing the geomertry of the black hole.
One has to introduce a certain assumption on the relation between
the field variables of the model and the tachyon wave functions
in the black hole background. This relation (somewhat modified
from the formulas proposed in [\MS,\DAS]) does not follow from
the model itself.

It turns out, however, the deformed model predicts the same scaling
exponents as ours and appears closely related to our
continuum discussions. First of all the free-energy of the
model has been computed [\JY] and behaves as $Z(M)\equiv M\log M$.
Thus the exponent $\gamma$ equals $1$ up to a logarithmic correction.
Also the characteristic features of our tachyon correlation functions
(22), (25) agree with those of the deformed model (for further
discussions on the deformed model see, [\D]).

In ref.[\WITB] Witten suggested a characterization of the integral
perturbations of the $c=1$ Liouville theory as the deformation of the
quadratic relation $a_1a_2-a_3a_4=0$ into a form
$$
a_1a_2-a_3a_4=\sum_{n}\varepsilon_n (a_1a_2)^n~,
\eqno\eq
$$
where $a_1=x\bar{x},a_2=y\bar{y},a_3=x\bar{y}$ and
$a_4=y\bar{x}$ and $x,y$ are the generators of the chiral ground ring.
Correspondence to the martrix model is given by $a_1=p+x,a_2=p-x$
where $x$ is the matrix eigenvalue and $p$ is its conjugate momentum.
In the case of the perturbation by the cosmological constant operator,
it is known [\KU] that the $n=0$ term contributes in the
right-hand-side
of (31) and the quadratic relation is deformed to $p^2-x^2=\mu$
which is exactly the equation for the fermi surface of the matrix
quantum mechanics. This result may be understood as the consequence of
the Liouville-momentum conservation: the ground ring elements $x,y$
have the form
$(bc,\partial\phi,\partial X)\exp(\pm i{1 \over \sqrt{2}}X)
\exp({1 \over \sqrt{2}}\phi)$ ($b,c$ are
the ghost fields) and hence the $a_1a_2$ carries a
Liouville-momentum $\sqrt{2}$. On the other
hand, the cosmological constant operator carries a momentum $-\sqrt{2}$.
Then the momentum of the operator $a_1a_2$ is cancelled after
perturbation and we obtain the
$n=0$ term in the right-hand-side of (31).

If we apply the same analysis in the case of the mass perturbation,
we obtain the $n=-1$ term in (31): the mass operator carries a
momentum $-2\sqrt{2}$ which converts the
momentum of $a_1a_2$ to $-\sqrt{2}$ under perturbation.
Thus eq.(31) is expected to behave as
$$
p^2-x^2={M \over p^2-x^2}~.
\eqno\eq
$$
If we ignore the $p^2$ piece in the denominator
of the right-hand-side, we obtain the Hamiltonian of the deformed
model.
Thus there is a good chance that the deformed matrix model is
equivalent to the mass-perturbed Liouville theory.

As we have remarked before, the black-hole mass term coincides
with the discrete state operator $W^-_{1,0}\bar{W}^-_{1,0}$ up to a
total derivative.
Since the cosmological constant operator
$W^-_{0,0}\bar{W}^-_{0,0}$ has also a similar structure, we may
consider
a general class of perturbations of the Liouville theory due to
negative discrete states $W^-_{j,0}\bar{W}^-_{j,0}, j=1,2,\cdots$.
There exists some indication that the operator
$W^-_{j,0}\bar{W}^-_{j,0}$ corresponds to a potential
$Tr\Phi^{-2j}$ in the matrix model: if we again use the
Liouville-momentum analysis, we find that the
$n=-j$
term is generated in the right-hand-side of (31) under the perturbation
by $W^-_{j,0}\bar{W}^-_{j,0}$. The $n=-j$ term corresponds to
the potential $Tr\Phi^{-2j}$.
It will be
extremely interesting to see if we can establish a definite relation
between
the perturbed Liouville theories and the deformed matrix models.

Materials of this paper have been presented at the workshop
"Quantum Aspects of the Black Hole" held at ITP, Santa Barbara,
June 21-26. I would like to thank participants of the meeting for
their discussions. In particular I am grateful to Prof. T. Yoneya for
detailed discussions on the deformed matrix model.
I am also grateful to Prof. C. Vafa for explaining a somewhat different
view on the two-dimensional black hole.
Research of T. Eguchi is partly supported by the Grant-in-Aid for
Scientific Research on Priority Area "Infinite Analysis".

\refout
\bye